\documentclass{elsarticle}

\usepackage{amsmath,amstext,amssymb}
\usepackage{psfrag,subfigure,xcolor,color}
\usepackage{graphicx}
\def\N{{\mathcal{N}}}

\def\k{\kappa_5}

\def\mt{{\tilde m}}
\def\a{\alpha}
\def\ra{\rightarrow}
\def\phit{\tilde{\phi}}

\def\del{\partial}

\newcommand{\dd}{\mathrm{d}}

\begin{document}

%%%%%%%%%%%%%%%%%%%%%%%%%%%%%%%% T I T L E P A G E (Phys Lett B)
\title{On Holographic p-wave Superfluids with Back-reaction}
\date{\today}

\cortext[cor1]{Corresponding Author}

\author[MPI]{Martin Ammon}
\ead{ammon@mppmu.mpg.de}
\author[MPI]{Johanna Erdmenger}    
\ead{jke@mppmu.mpg.de}
\author[MPI]{Viviane Grass\corref{cor1}} 
\ead{grass@mppmu.mpg.de}
\author[MPI]{Patrick Kerner}
\ead{pkerner@mppmu.mpg.de}
\author[MPI]{Andy O'Bannon}
\ead{ahob@mppmu.mpg.de}

\address[MPI]{Max-Planck-Institut f\"ur Physik (Werner-Heisenberg-Institut)\\
  F\"ohringer Ring 6, 80805 M\"unchen, Germany}

%\preprint{MPP-2009-215}

%%%%%%%%%%%%%%%%%%%%%%%%%%%%%%%% A B S T R A C T
\begin{abstract}
\noindent 
We numerically construct asymptotically Anti-de Sitter charged black hole
solutions of $(4+1)$-dimensional $SU(2)$ Einstein-Yang-Mills theory that, for
sufficiently low temperature, develop vector hair.  Via gauge-gravity duality,
these solutions describe a strongly-coupled conformal field theory at finite
temperature and density that undergoes a phase transition to a superfluid
state with spontaneously broken rotational symmetry (a p-wave superfluid
state). The bulk theory has a single free parameter, the ratio of the
five-dimensional gravitational constant to the Yang-Mills coupling, which we
denote as $\a$. Previous analyses have shown that in the so-called probe
limit, where $\a$ goes to zero and hence the gauge fields are ignored in
Einstein's equation, the transition to the superfluid state is second order.
We construct fully back-reacted solutions, where $\a$ is finite and the gauge
fields are included in Einstein's equation, and find that for values of $\a$
above a critical value $\a_c = 0.365 \pm 0.001$ in units of the AdS radius, the transition becomes first
order. 
\end{abstract}

\begin{keyword}
  Gauge/gravity duality, Black Holes, Phase transitions
  \PACS 11.25.Tq, 04.70.Bw, 05.70.Fh
\end{keyword}

\maketitle

%%%%%%%%%%%%%%%%%%%%%%%%%%%%%%%%%%%%%%%%%%%%%%%%%%%%%%%%%%%%%%%%%%%%%%%%%%%%%%%%%%%%%
\section{Introduction}
%%%%%%%%%%%%%%%%%%%%%%%%%%%%%%%%%%%%%%%%%%%%%%%%%%%%%%%%%%%%%%%%%%%%%%%%%%%%%%%%%%%%%

The Anti-de Sitter/Conformal Field Theory correspondence (AdS/CFT) \cite{Maldacena:1997re} provides a novel method for studying strongly-coupled systems at finite density. As such, it may have useful applications in condensed matter physics, especially for studying scale-invariant strongly-coupled systems, for example, low-temperature systems near quantum criticality (see for example refs.\cite{Herzog:2009xv,Hartnoll:2009sz,McGreevy:2009xe} and references therein). Such systems are not purely theoretical: the (thermo)dynamics of some high-$T_c$ superconductors may be controlled by a quantum critical point.

AdS/CFT is a holographic duality: it equates a weakly-coupled theory of
gravity in $(d+1)$-dimensional AdS space with a strongly-coupled
$d$-dimensional CFT ``living'' at the AdS boundary. CFT states with finite temperature 
are dual to black hole geometries, where
the Hawking temperature of the black hole is identified with the temperature
in the CFT \cite{Witten:1998zw}. The current of a global $U(1)$ symmetry in the CFT will be dual to
a $U(1)$ gauge field in AdS space. AdS/CFT thus allows us to compute
observables in a strongly-coupled CFT, in states with finite temperature and
density, by studying asymptotically AdS charged black holes. By now AdS/CFT
can model many basic phenomena in condensed matter physics, such as the
quantum Hall effect \cite{Davis:2008nv}, non-relativistic scale-invariance \cite{Son:2008ye, Balasubramanian:2008dm}, and
Fermi surfaces \cite{Rey:2008zz,Liu:2009dm, Cubrovic:2009ye}.  

AdS/CFT can also describe phase transitions to superfluid states, that is,
phase transitions in which a sufficiently large $U(1)$ charge density triggers
spontaneous breaking of the $U(1)$ symmetry: an operator charged under the
$U(1)$ acquires a nonzero expectation value
\cite{Gubser:2008px,Hartnoll:2008vx,Hartnoll:2008kx}. We will refer to this as the operator
``condensing.'' The simplest bulk action that can describe such a transition
is Einstein-Maxwell theory coupled to a charged scalar. In the bulk, a charged
black hole develops scalar hair. In the CFT, a charged scalar operator
condenses.  

A simple bulk action has one great virtue, namely a kind of universality: the
results may be true for many different dual CFT's, independent of the details
of their dynamics. For the Einstein-Maxwell-scalar case, a fruitful
exercise is to study various functional forms for the scalar potential and to
scan through values of couplings in that potential \cite{Franco:2009yz,Franco:2009if,Aprile:2009ai}. Generally speaking,
scanning through values of these parameters corresponds to scanning through
many different dual CFT's. As shown in refs.
\cite{Franco:2009yz,Franco:2009if,Aprile:2009ai}, such changes can have a dramatic effect,
for example, the phase transition can change from second to first order. 

AdS/CFT can also describe superfluid states in which the condensing operator
is a vector and hence rotational symmetry is broken, that is, p-wave
superfluid states \cite{Gubser:2008wv,Herzog:2009ci}. Here the
CFT has a global $SU(2)$ symmetry and hence three conserved currents
$J^{\mu}_a$, where $a=1,2,3$ label the generators of $SU(2)$. For a
sufficiently large charge density for some $U(1)$ subgroup of $SU(2)$, say a
sufficiently large $\langle J^t_3\rangle$, holographic calculations reveal
that, of the known available states, those with lowest free energy have a
nonzero $\langle J^x_1 \rangle$. Not only is the $U(1)$ broken, but spatial
rotational symmetry is also broken to some subgroup. 

On the AdS side, a simple bulk action that can describe such a transition is
Einstein-Yang-Mills theory with gauge group $SU(2)$. CFT states with nonzero
$\langle J^t_3\rangle$ are dual to black hole solutions with nonzero vector field
$A_t^3(r)$ in the bulk, where $r$ is the radial coordinate of AdS space.
States with nonzero $\langle J^x_1\rangle$ are dual to black hole solutions
with a nontrivial $A_x^1(r)$. The superfluid phase transition is thus dual to
charged AdS black holes developing vector hair. A string theory realization for
this model is given in refs. \cite{Ammon:2008fc,Basu:2008bh,Ammon:2009fe,Peeters:2009sr}.

Unlike the Einstein-Maxwell-scalar case, $SU(2)$ Einstein-Yang-Mills theory has only a \textit{single} free
parameter, $\alpha\equiv \k / \hat{g}$, where $\k$ is the gravitational constant (we
will work in $(4+1)$ dimensions, hence the subscript) and $\hat{g}$ is the
Yang-Mills coupling. The Yang-Mills source terms on the right-hand-side of
Einstein's equation are proportional to $\alpha^2$. To date, most analyses of the holographic p-wave superfluid transition have
employed the so-called probe limit, which consists of taking $\alpha \ra 0$ so that the gauge fields have no effect on the geometry,
which becomes simply AdS-Schwarzschild. The probe limit was sufficient to show
that a superfluid phase transition occurs and is second order. 

Our goal is to study the effect of finite $\alpha$, that is, to study the
back-reaction of the gauge fields on the metric. We will work with
$(4+1)$-dimensional $SU(2)$ Einstein-Yang-Mills theory, with finite $\alpha$.
We will numerically construct asymptotically AdS charged black hole solutions
with vector hair (for similar studies see refs. \cite{Gubser:2008zu,Basu:2009vv,Manvelyan:2008sv}). 
Our principal result is that for a sufficiently large value
of $\alpha$ the phase transition becomes first order. More specifically, we
find a critical value $\alpha_c=0.365\pm0.001$ in units of the AdS radius, such that the transition is
second order when $\alpha<\alpha_c$ and first order when
$\alpha>\alpha_c$.

We can provide some intuition for what increasing $\alpha$ means, in CFT terms, as follows. Generically, in AdS/CFT $1/\k^2 \propto c$, where $c$ is the central charge of the CFT \cite{Henningson:1998gx,Balasubramanian:1999re}, which, roughly speaking, counts the total number of degrees of freedom in the CFT. Correlation functions involving the $SU(2)$ current will generically be proportional to $1/\hat{g}^2$ \cite{Herzog:2009xv,Hartnoll:2009sz,McGreevy:2009xe}. We may (again roughly) think of $1/\hat{g}^2$ as counting the number of degrees of freedom in the CFT that carry $SU(2)$ charge. Intuitively, then, in the CFT increasing $\alpha$ means increasing the ratio of charged degrees of freedom to total degrees of freedom.

The paper is organized as follows. In section \ref{setup} we write the action of our
model and discuss our ansatz for the bulk fields. In 
section \ref{thermo} we describe how to extract thermodynamic information from our
solutions. In section \ref{phasetrans} we present numerical results demonstrating that
increasing $\alpha$ changes the order of the phase transition. We conclude
with some discussion and suggestions for future research in section \ref{discuss}.  

%%%%%%%%%%%%%%%%%%%%%%%%%%%%%%%%%%%%%%%%%%%%%%%%%%%%%%%%%%%%%%%%%%%%%%%%%%%%%%%%%%%%%

\section{Holographic Setup}\label{setup}

%%%%%%%%%%%%%%%%%%%%%%%%%%%%%%%%%%%%%%%%%%%%%%%%%%%%%%%%%%%%%%%%%%%%%%%%%%%%%%%%%%%%%

We consider $SU(2)$ Einstein-Yang-Mills theory in $(4+1)$-dimensional asymptotically AdS space. The action is
\begin{equation}
\label{eq:action}
S = \int\!\dd^5x\,\sqrt{-g} \, \left [ \frac{1}{2\k^2} \left( R -\Lambda\right) - \frac{1}{4\hat g^2} \, F^a_{\mu\nu} F^{a\mu \nu} \right] + S_{bdy}\,, 
\end{equation}
where $\k$ is the five-dimensional gravitational constant, $\Lambda = - \frac{12}{L^2}$ is the cosmological constant, with $L$ being the AdS radius, and $\hat g$ is the Yang-Mills coupling constant. The $SU(2)$ field strength $F^a_{\mu\nu}$ is
\begin{equation}
F^a_{\mu\nu}=\del_\mu A^a_\nu -\del_\nu A^a_\mu + \epsilon^{abc}A^b_\mu A^c_\nu \,,
\end{equation}
where $\mu, \nu = \{t,r,x,y,z\}$, with $r$ being the AdS radial coordinate, and $\epsilon^{abc}$ is the totally antisymmetric tensor with $\epsilon^{123}=+1$. The $A^a_\mu$ are the components of the matrix-valued gauge field, $A=A^a_\mu\tau^a dx^\mu$,  where the $\tau^a$ are the $SU(2)$ generators, which are related to the Pauli matrices by $\tau^a=\sigma^a/2i$.  $S_{bdy}$ includes boundary terms that do not affect the equations of motion, namely the Gibbons-Hawking boundary term as well as counterterms required for the on-shell action to be finite. We will write $S_{bdy}$ explicitly in section \ref{thermo}. 

The Einstein and Yang-Mills equations derived from the above action are
\begin{align}
\label{eq:einsteinEOM}
R_{\mu \nu}+\frac{4}{L^2}g_{\mu \nu}&=\k^2\left(T_{\mu\nu}-\frac{1}{3}{T_{\rho}}^{\rho}g_{\mu\nu}\right)\,, \\
\label{eq:YangMillsEOM}
\nabla_\mu F^{a\mu\nu}&=-\epsilon^{abc}A^b_\mu F^{c\mu\nu} \,,
\end{align}
where the Yang-Mills stress-energy tensor $T_{\mu\nu}$ is
\begin{equation}
\label{eq:energymomentumtensor}
T_{\mu \nu}=\frac{1}{\hat{g}^2}{\rm tr}\left(F_{\rho\mu}{F^{\rho}}_{\nu}-\frac{1}{4}g_{\mu\nu} F_{\rho\sigma}F^{\rho\sigma}\right)\,.
\end{equation}

Following ref. \cite{Gubser:2008wv}, to construct charged black hole solutions with vector hair we choose a gauge field ansatz
\begin{equation}
\label{eq:gaugefieldansatz}
A=\phi(r)\tau^3\dd t+w(r)\tau^1\dd x\,.
\end{equation}
The motivation for this ansatz is as follows. In the field theory we will
introduce a chemical potential for the $U(1)$ symmetry generated by $\tau^3$.
We will denote this $U(1)$ as $U(1)_3$. The bulk operator dual to the $U(1)_3$
density is $A^3_t$, hence we include $A^3_t(r) \equiv \phi(r)$ in our ansatz.
We want to allow for states with a nonzero $\langle J^x_1\rangle$, so in
addition we introduce $A^1_x(r) \equiv w(r)$. With this ansatz for the gauge
field, the Yang-Mills stress-energy tensor in eq.~\eqref{eq:energymomentumtensor} is
diagonal. Solutions with nonzero $w(r)$
will preserve only an $SO(2)$ subgroup of the $SO(3)$ rotational symmetry, so
our metric ansatz will respect only $SO(2)$. Furthermore, given that the
Yang-Mills stress-energy tensor is diagonal, a diagonal metric is consistent. We will also pattern our metric ansatz after the ones used in ref.
\cite{Manvelyan:2008sv} since these tame singular points in the equations of
motion. Our metric ansatz is  
\begin{equation}\label{eq:metricansatz}
\dd s^2 = -N(r)\sigma(r)^2\dd t^2 + \frac{1}{N(r)}\dd r^2 +r^2 f(r)^{-4}\dd x^2 + r^2f(r)^2\left(\dd y^2 + \dd z^2\right)\,,
\end{equation}
with $N(r)=-\frac{2m(r)}{r^2}+\frac{r^2}{L^2}$. For our black hole solutions we will denote the position of the horizon as $r_h$. The AdS boundary will be at $r\rightarrow\infty$. 

Inserting our ansatz into the Einstein and Yang-Mills equations yields five equations of motion for  $m(r),\,\sigma(r),\,f(r),\,\phi(r),\,w(r)$ and one constraint equation from the $rr$ component of the Einstein equations. The dynamical equations can be recast as (prime denotes $\frac{\partial}{\partial r}$)

\begin{equation}\label{eom}
 \begin{split}
m' &= \frac{\alpha^2 r f^4 w^2 \phi^2}{6 N \sigma^2} + \frac{\alpha^2 r^3 {\phi'}^2}{6 \sigma^2} + N\left(\frac{r^3{f'}^2}{f^2} + \frac{\alpha^2}{6} r f^4 {w'}^2\right) \,, \\ \sigma' &= \frac{\alpha^2 f^4 w^2 \phi^2}{3 r N^2 \sigma} + \sigma\left(\frac{2 r {f'}^2}{f^2} + \frac{\alpha^2 f^4 {w'}^2}{3 r}\right) \,, \\ f'' &= -\frac{\alpha^2 f^5 w^2 \phi^2}{3 r^2 N^2 \sigma^2} + \frac{\alpha^2 f^5 {w'}^2}{3 r^2} - f'\left(\frac{3}{r} - \frac{f'}{f} + \frac{N'}{N} +\frac{\sigma'}{\sigma}\right) \,, \\ \phi'' &= \frac{f^4 w^2 \phi}{r^2 N} - \phi'\left(\frac{3}{r} + \frac{\sigma'}{\sigma}\right) \,, \\ w'' &= -\frac{w \phi^2}{N^2 \sigma^2} - w'\left( \frac{1}{r} + \frac{4 f'}{f} + \frac{N'}{N} + \frac{\sigma'}{\sigma} \right).  \,
\end{split}
\end{equation}

The equations of motion are invariant under four scaling transformations (invariant quantities are not shown),
\begin{eqnarray}
(I) & \sigma\rightarrow \lambda\sigma, \qquad \phi\rightarrow \lambda\phi,& \nonumber \\ (II)& f\rightarrow \lambda f, \qquad w\rightarrow \lambda^{-2} w,& \nonumber \\  (III) & r\rightarrow \lambda r\,, \quad m\rightarrow \lambda^4 m \,, \quad w\rightarrow \lambda w \,, \quad \phi\rightarrow \lambda\phi, \, & \nonumber \\  (IV) & r\rightarrow \lambda r\,,\quad m\rightarrow \lambda^2 m\,, \quad L\rightarrow \lambda L\,,\quad \phi\rightarrow \frac{\phi}{\lambda}\,, \quad \alpha\rightarrow \lambda \alpha, & \nonumber
\end{eqnarray}
where in each case $\lambda$ is some real positive number. Using (I) and (II) we can set the boundary values of both $\sigma(r)$ and $f(r)$ to one, so that the metric will be asymptotically AdS. We are free to use (III) to set $r_h$ to be one, but we will retain $r_h$ as a bookkeeping device. We will use (IV) to set the AdS radius $L$ to one.

A known analytic solution of the equations of motion is an asymptotically AdS Reissner-Nordstr\"om black hole, which has $\phi(r)=\mu - q/r^2$, $w(r)=0$, $\sigma(r)=f(r)=1$, and $N(r)= \left(r^2 - \frac{2m_0}{r^2} + \frac{2\alpha^2 q^2}{3 r^4}\right)$, where $m_0=\frac{r_h^4}{2}+\frac{\alpha^2 q^2}{3r_h^2}$ and $q= \mu r^2_h$. Here $\mu$ is the value of $\phi(r)$ at the boundary, which in CFT terms is the $U(1)_3$ chemical potential.

To find solutions with nonzero $w(r)$ we resort to numerics. We will solve the equations of motion using a shooting method. We will vary the values of functions at the horizon until we find solutions with suitable values at the AdS boundary. We thus need the asymptotic form of solutions both near the horizon $r=r_h$ and near the boundary $r=\infty$.

Near the horizon, we define $\epsilon_h\equiv\frac{r}{r_h}-1\ll 1$ and then expand every function in powers of $\epsilon_h$ with some constant coefficients. Two of these we can fix as follows. We determine $r_h$ by the condition $N(r_h)=0$, which gives that $m(r_h)=r_h^4/2$. Additionally, we must impose $A^3_t(r_h)=\phi(r_h)=0$ for $A$ to be well-defined as a one-form (see for example ref. \cite{Kobayashi:2006sb}). The equations of motion then impose relations among all the coefficients. A straightforward exercise shows that only four coefficients are independent,
\begin{equation}
\left\{\phi^h_1, \sigma^h_0, f^h_0, w^h_0\right\} \,,
\end{equation}
where the subscript denotes the order of $\epsilon_h$ (so $\sigma^h_0$ is the value of $\sigma(r)$ at the horizon, etc.). All other near-horizon coefficients are determined in terms of these four.

Near the boundary $r=\infty$ we define $\epsilon_b\equiv \left(\frac{r_h}{r}\right)^2\ll1$ and then expand every function in powers of $\epsilon_b$ with some constant coefficients. The equations of motion again impose relations among the coefficients. The independent coefficients are
\begin{equation}\label{coeffb}
\left\{m^b_0, \mu, \phi^b_1, w^b_1, f^b_2\right\} \,,
\end{equation}
where here the subscript denotes the power of $\epsilon_b$. All other near-boundary coefficients are determined in terms of these.

We used scaling symmetries to set $\sigma_0^b = f_0^b=1$. Our solutions will also have $w_0^b=0$ since we do not want to source the operator $J^x_1$ in the CFT ($U(1)_3$ will be \textit{spontaneously} broken). In our shooting method we choose a value of $\mu$ and then vary the four independent near-horizon coefficients until we find a solution which produces the desired value of $\mu$ and has $\sigma_0^b = f_0^b=1$ and $w_0^b=0$.

In what follows we will often work with dimensionless coefficients by scaling
out factors of $r_h$. We thus define the dimensionless functions
$\mt(r)\equiv m(r)/r_h^4$, $\tilde\phi(r)\equiv \phi(r)/r_h$ and $\tilde w(r)\equiv w(r)/r_h$,
while $f(r)$ and $\sigma(r)$ are already dimensionless.

%%%%%%%%%%%%%%%%%%%%%%%%%%%%%%%%%%%%%%%%%%%%%%%%%%%%%%%%%%%%%%%%%%%%%%%%%%%%%%%%%%%%%
\section{Thermodynamics}\label{thermo}
%%%%%%%%%%%%%%%%%%%%%%%%%%%%%%%%%%%%%%%%%%%%%%%%%%%%%%%%%%%%%%%%%%%%%%%%%%%%%%%%%%%%%

Next we will describe how to extract thermodynamic information from our solutions. Our solutions describe thermal equilibrium states in the dual CFT. We will work in the grand canonical ensemble, with fixed chemical potential $\mu$.

We can obtain the temperature and entropy from horizon data. The temperature $T$ is given by the Hawking temperature of the black hole,
\begin{equation}
  \label{eq:temperature}
  T=\frac{\kappa}{2\pi}=\frac{\sigma^h_0}{12\pi}\left(12-\alpha^2 \frac{{(\phit^h_1)}^2}{{\sigma^h_0}^2}\right)\,r_h\,.
\end{equation}
Here $\kappa=\left . \sqrt{\del_\mu \xi \del^\mu \xi} \right |_{r_h}$ is the surface gravity of the black hole, with $\xi$ being the norm of the timelike Killing vector, and in the second equality we write $T$ in terms of near-horizon coefficients. In what follows we will often convert from $r_h$ to $T$ simply by inverting the above equation. The entropy $S$ is given by the Bekenstein-Hawking entropy of the black hole,
\begin{equation}
  \label{eq:entropy}
  S=\frac{2\pi}{\k^2}A_h=\frac{2\pi V}{\k^2}r_h^3=  \frac{2\pi^4}{\k^2}VT^3
  \frac{12^3{\sigma_0^h}^3}{\left(12{\sigma_0^h}^2-{(\phit_1^h)}^2\alpha^2\right)^3}\,,
\end{equation}
where $A_h$ denotes the area of the horizon and $V = \int\!\dd^3x$.

The central quantity in the grand canonical ensemble is the grand potential $\Omega$. In AdS/CFT we identify $\Omega$ with $T$ times the on-shell bulk action in Euclidean signature. We thus analytically continue to Euclidean signature and compactify the time direction with period $1/T$. We denote the Euclidean bulk action as $I$ and $I_{\text{on-shell}}$ as its on-shell value (and similarly for other on-shell quantities). Our solutions will always be static, hence $I_{\text{on-shell}}$ will always include an integration over the time direction, producing a factor of $1/T$. To simplify expressions, we will define $I \equiv \tilde{I}/T$.  Starting now, we will refer to $\tilde{I}$ as the action. $\tilde{I}$ includes a bulk term, a Gibbons-Hawking boundary term, and counterterms,
\begin{equation}
\label{eq:renomaction}
  \tilde{I}=\tilde{I}_{\text{bulk}}+\tilde{I}_{\text{GH}}+\tilde{I}_{\text{CT}}\,.
\end{equation}
$\tilde{I}_{\text{bulk}}^{\text{on-shell}}$ and $\tilde{I}_{GH}^{\text{on-shell}}$ exhibit divergences, which are canceled by the counterterms in $\tilde{I}_{\text{CT}}$. To regulate these divergences we introduce a hypersurface $r=r_{bdy}$ with some large but finite $r_{bdy}$. We will always ultimately remove the regulator by taking $r_{bdy}\rightarrow\infty$. Using the equations of motion, for our ansatz $\tilde{I}_{\text{bulk}}^{\text{on-shell}}$ is
\begin{equation}
 \tilde{I}_{\text{bulk}}^{\text{on-shell}}=\frac{V}{\k^2} \,\frac{1}{2f^2}r N \sigma (r^2 f^2)'  \Big|_{r=r_{\text{bdy}}} \,.
\end{equation}
For our ansatz, the Euclidean Gibbons-Hawking term is
\begin{equation}
  \label{eq:GHterm}
  \tilde{I}_{\text{GH}}^{\text{on-shell}}=-\frac{1}{\k^2}\int\!\dd^3x\sqrt{\gamma}\,\nabla_\mu n^\mu=-\frac{V}{\k^2}N\sigma r^3\left(\frac{N'}{2N}+\frac{\sigma'}{\sigma}+\frac3r\right)\Big|_{r=r_{\text{bdy}}}\,,
\end{equation}
where $\gamma$ is the induced metric on the $r=r_{bdy}$ hypersurface and $n_\mu\dd x^\mu=1/\sqrt{N(r)}\,\dd r$ is the outward-pointing normal vector. The only divergence in $\tilde{I}_{\text{bulk}}^{\text{on-shell}} + \tilde{I}_{\text{GH}}^{\text{on-shell}}$ comes from the infinite volume of the asymptotically AdS space, hence, for our ansatz, the only nontrivial counterterm is
\begin{equation}
  \label{eq:counterterms}
  \tilde{I}_{\text{CT}}^{\text{on-shell}}=\frac{3}{\k^2}\int\!\dd^3x\sqrt{\gamma}=\frac{3 V}{\k^2}r^3\sqrt{N}\sigma\Big|_{r=r_{\text{bdy}}}\,.
\end{equation}
Finally, $\Omega$ is related to the on-shell action, $\tilde{I}_{\text{on-shell}}$, as
\begin{equation}
\Omega = \lim_{r_{bdy}\rightarrow\infty} \tilde{I}_{\text{on-shell}}.
\end{equation}

The chemical potential $\mu$ is simply the boundary value of $A^3_t(r) = \phi(r)$. The charge density $\langle J^t_3\rangle$ of the dual field theory can be extracted from $\tilde{I}_{\text{on-shell}}$ by
\begin{eqnarray}
 \langle J^t_3\rangle= \frac{1}{V} \, \lim_{r_{bdy}\rightarrow \infty} \frac{\delta \tilde{I}_{\text{on-shell}}}{\delta A_t^3(r_{bdy})} =-\frac{2\pi^3\alpha^2}{\k^2}T^3
 \frac{12^3{\sigma_0^h}^3}{\left(12{\sigma_0^h}^2-{(\phit_1^h)}^2\alpha^2\right)^3} \, \tilde{\phi}^b_1 \,.
\end{eqnarray}
Similarly, the current density $\langle J^x_1\rangle$ is
\begin{eqnarray}
 \langle J^x_1\rangle= \frac{1}{V} \, \lim_{r_{bdy}\rightarrow \infty} \frac{\delta \tilde{I}_{\text{on-shell}}}{\delta A_x^1(r_{bdy})} =+\frac{2\pi^3\alpha^2}{\k^2}T^3
 \frac{12^3{\sigma_0^h}^3}{\left(12{\sigma_0^h}^2-{(\phit_1^h)}^2\alpha^2\right)^3} \, \tilde{w}^b_1 \,.
\end{eqnarray}

The expectation value of the stress-energy tensor of the CFT is \cite{Balasubramanian:1999re, deHaro:2000xn}
\begin{equation}
  \label{eq:energymombdy}
  \langle T_{ij}\rangle=\lim_{r_{bdy}\rightarrow \infty} \frac{2}{\sqrt{\gamma}}\frac{\delta \tilde{I}_{\text{on-shell}}}{\delta \gamma^{ij}}= \lim_{r_{bdy}\rightarrow \infty} \left [ \frac{r^2}{\k^2}
  \left(-K_{ij}+{K^l}_l\gamma_{ij}-3\,\gamma_{ij}\right) \right ]_{r=r_{\text{bdy}}} \,,
\end{equation}
where $i, j, l = \{t,x,y,z\}$ and $K_{ij}= \frac{1}{2} \sqrt{N(r)} \, \partial_r \gamma_{ij}$ is the extrinsic curvature. We find
\begin{equation}
\label{eq:cftstressenergytensor}
 \begin{split}
\langle T_{tt} \rangle&=3\frac{\pi^4}{\k^2}VT^4\frac{12^4{\sigma_0^h}^4}{\left(12{\sigma_0^h}^2-{(\phit_1^h)}^2\alpha^2\right)^4}\,\mt^b_0 \,,\\
  \langle T_{xx} \rangle&= \frac{\pi^4}{\k^2}VT^4\frac{12^4{\sigma^h_0}^4}{\left(12{\sigma^h_0}^2-{(\phit_1^h)}^2\alpha^2\right)^4}\left(\mt^b_0-8f_2^b\right)\,,\\
  \langle T_{yy} \rangle = \langle T_{zz} \rangle&= \frac{\pi^4}{\k^2}VT^4\frac{12^4{\sigma^h_0}^4}{\left(12{\sigma^h_0}^2-{(\phit_1^h)}^2\alpha^2\right)^4}\left(\mt^b_0+4f_2^b\right)\,.
 \end{split}
\end{equation}
Notice that $\langle T_{tx} \rangle = \langle T_{ty} \rangle = \langle T_{tz}
\rangle = 0$. Even in phases where the current $\langle J_1^x \rangle$ is
nonzero, the fluid will have zero net momentum. Indeed, this result is
guaranteed by our ansatz for the gauge field which implies a diagonal
Yang-Mills stress-energy tensor and a diagonal metric (the spacetime is static).

For $\mt^b_0=\frac12+\frac{\alpha^2 \tilde\mu^2}{3}$, $\sigma_0^h=1$, ${\phit_1^h}=2\tilde\mu$, $f_2^b=0$, and $\tilde\phi^b_1=-\tilde\mu$ we recover the correct thermodynamic properties of the Reissner-Nordstr\"om black hole, which preserves the $SO(3)$ rotational symmetry. For example, we find that $\langle T_{xx} \rangle=\langle T_{yy} \rangle = \langle T_{zz} \rangle$ and $\Omega=-\langle T_{yy}\rangle$. For solutions with nonzero $\langle J_1^x\rangle$, the $SO(3)$  is broken to $SO(2)$. In these cases, we find that $\langle T_{xx} \rangle \neq \langle T_{yy} \rangle = \langle T_{zz} \rangle$. Just using the equations above, we also find $\Omega = - \langle T_{yy} \rangle$. In the superfluid phase, both the nonzero $\langle J^x_1 \rangle$ and the stress-energy tensor indicate breaking of $SO(3)$. 

Tracelessness of the stress-energy tensor (in Lorentzian signature) implies $\langle T_{tt} \rangle= \langle T_{xx} \rangle + \langle T_{yy} \rangle + \langle T_{zz} \rangle$, which is indeed true for eq. (\ref{eq:cftstressenergytensor}), so in the dual CFT we always have a conformal fluid. The only physical parameter in the CFT is thus the ratio $\mu/T$.

%%%%%%%%%%%%%%%%%%%%%%%%%%%%%%%%%%%%%%%%%%%%%%%%%%%%%%%%%%%%%%%%%%%%%%%%%%%%%%%%%%%%%%%%%%%%%%
\section{Phase Transitions}\label{phasetrans}
%%%%%%%%%%%%%%%%%%%%%%%%%%%%%%%%%%%%%%%%%%%%%%%%%%%%%%%%%%%%%%%%%%%%%%%%%%%%%%%%%%%%%%%%%%%%%%

\begin{figure}
\centering
\subfigure[]{\includegraphics[width=0.45\linewidth]{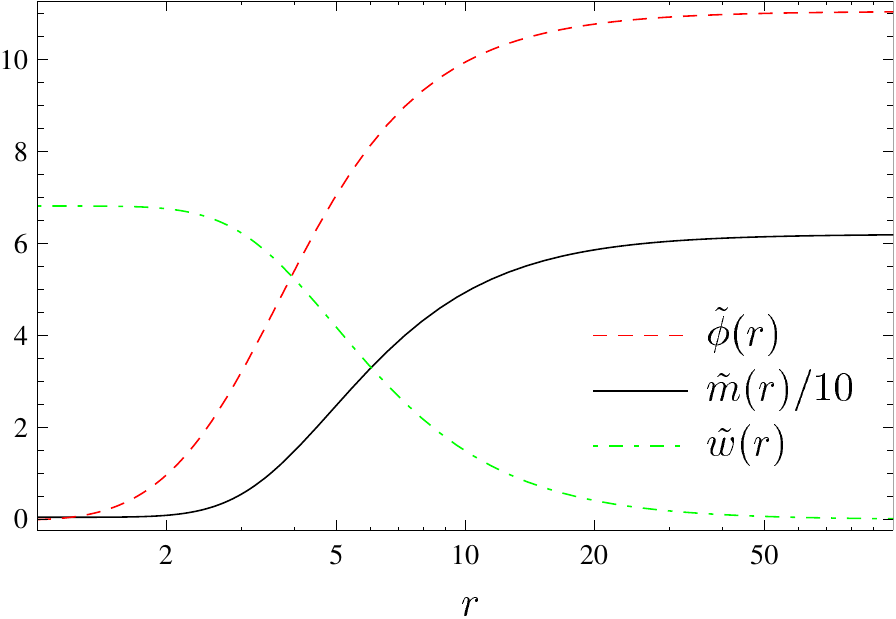}}
\hfill
\subfigure[]{\includegraphics[width=0.45\linewidth]{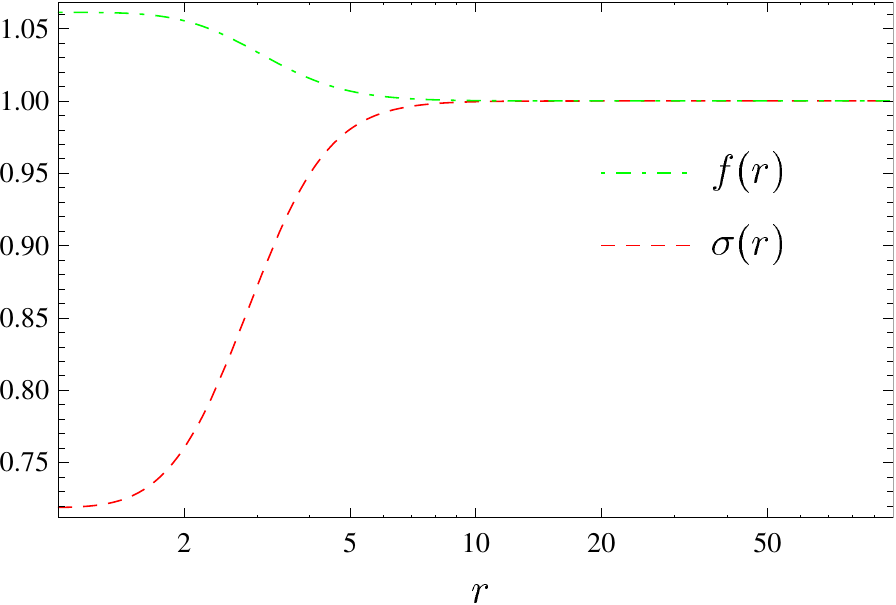}}
\caption{(a) The dimensionless gauge field components $\phit(r)$ (red dashed) and $\tilde{w}(r)$ (green dot-dashed) and the dimensionless metric function $\mt(r)$, scaled down by a factor of $10$, (black solid) versus the AdS radial coordinate $r$ for $\alpha=0.316$ at $T\approx 0.45 T_c$. (b) The dimensionless metric functions $\sigma(r)$ (red dashed) and $f(r)$ (green dot-dashed) versus the AdS radial coordinate $r$ for $\alpha=0.316$ at $T\approx 0.45 T_c$.}
\label{fig:metricgaugealpha0.1}
\end{figure}

In this section we present our numerical results. We scanned through values of $\alpha$ from $\alpha=0.032$ to $\alpha=0.548$. Typical solutions for the metric and gauge field functions appear in Figure \ref{fig:metricgaugealpha0.1}. The solutions for other values of $\alpha$ are qualitatively similar. Notice that all boundary conditions are met: at the horizon $\phit(r)$ vanishes, and at the boundary$f_0^b=\sigma_0^b=1$ and $\tilde{w}_0^b=0$.

For every value of $\alpha$ that we use, we find Reissner-Nordstr\"om solutions for all temperatures, and for sufficiently low temperatures we always find additional solutions, with nonzero $w(r)$, that are thermodynamically preferred to the Reissner-Nordstr\"om solution. In other words, for every value of $\alpha$ that we use, we find a phase transition, at some temperature $T_c$, in which a charged black hole grows vector hair, which in the CFT is a p-wave superfluid phase transition. Our numerical results show that the phase transition is second order for $\alpha<\alpha_c$ and first order for $\alpha>\alpha_c$ where $\alpha_c\approx 0.365\pm0.001$.

\begin{figure}
  \centering
  \includegraphics[width=0.7\linewidth]{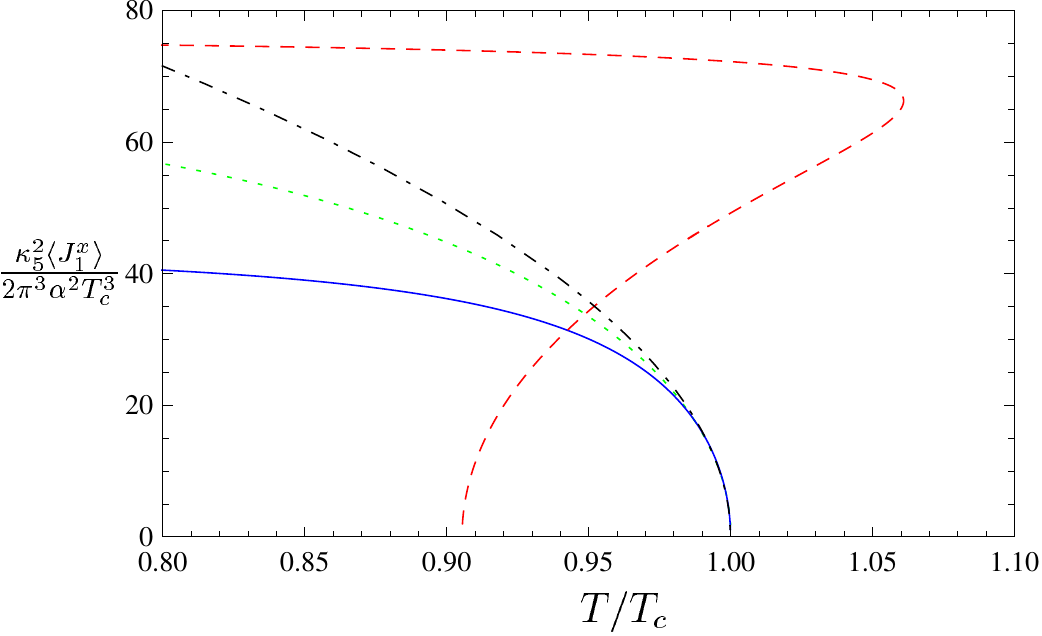}
  \caption{The order parameter $\langle J^x_1\rangle$, multiplied by
    $\k^2/(2\pi^3\alpha^2T_c^3)$, versus the rescaled temperature $T/T_c$ for
    different $\alpha$: $\alpha=0.032<\alpha_c$ (green dotted),
    $\alpha=0.316<\alpha_c$ (blue solid) and $\alpha=0.447>\alpha_c$ (red dashed).  The black dot-dashed curve is the function
    $a(1-T/T_c)^{1/2}$ with $a=160$. The green dotted curve is scaled up by a factor
    of $8$ while the red dashed curve is scaled down by a factor of $5$ such
    that $a$, which depends on $\alpha$, coincides for the green dotted and blue
    solid curves. If we
    decrease $T$ toward $T_c$, entering the figure from the right, we see that
    the blue solid and the green dotted curves rise continuously and
    monotonically from zero at $T=T_c$, signaling a second-order phase
    transition. The close agreement with the black dot-dashed curve suggests that these grow from zero as $\left(1-T/T_c\right)^{1/2}$. In the $\alpha=0.447$ case, the red dashed curve becomes
    multi-valued at $T=1.061\,T_c$. In this case, at
    $T=T_c$, the value of $\k^2\langle J^x_1 \rangle/(2\pi^3\alpha^2T_c^3)$
    jumps from zero to the upper part of the red dashed curve, signaling a first-order transition.}
  \label{fig:cond}
\end{figure}

\begin{figure}
  \centering
  \subfigure[]{\includegraphics[width=0.47\linewidth]{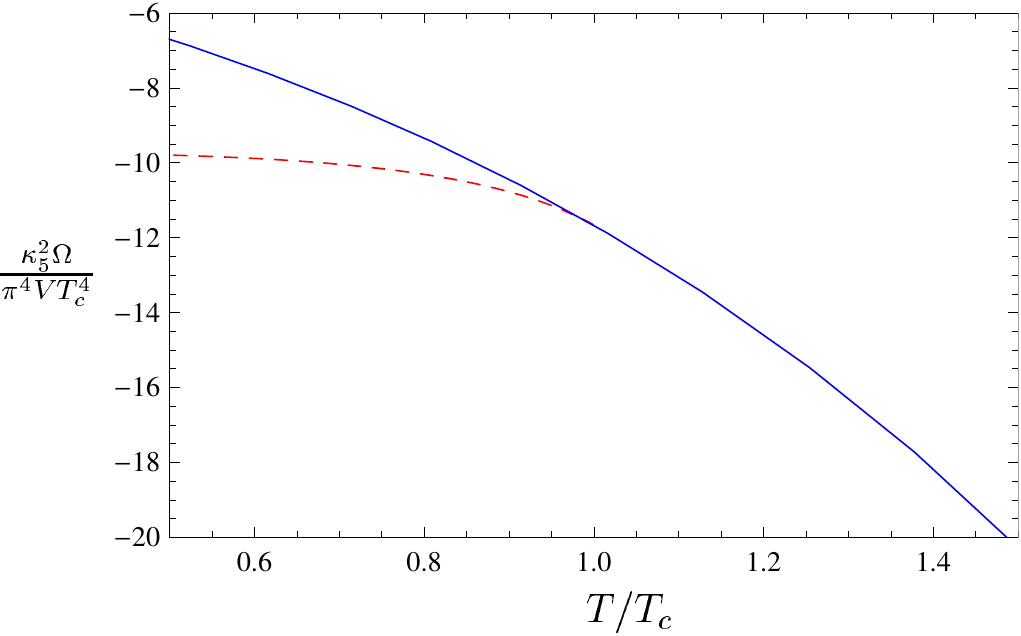}}
  \hfill
  \subfigure[]{\includegraphics[width=0.47\linewidth]{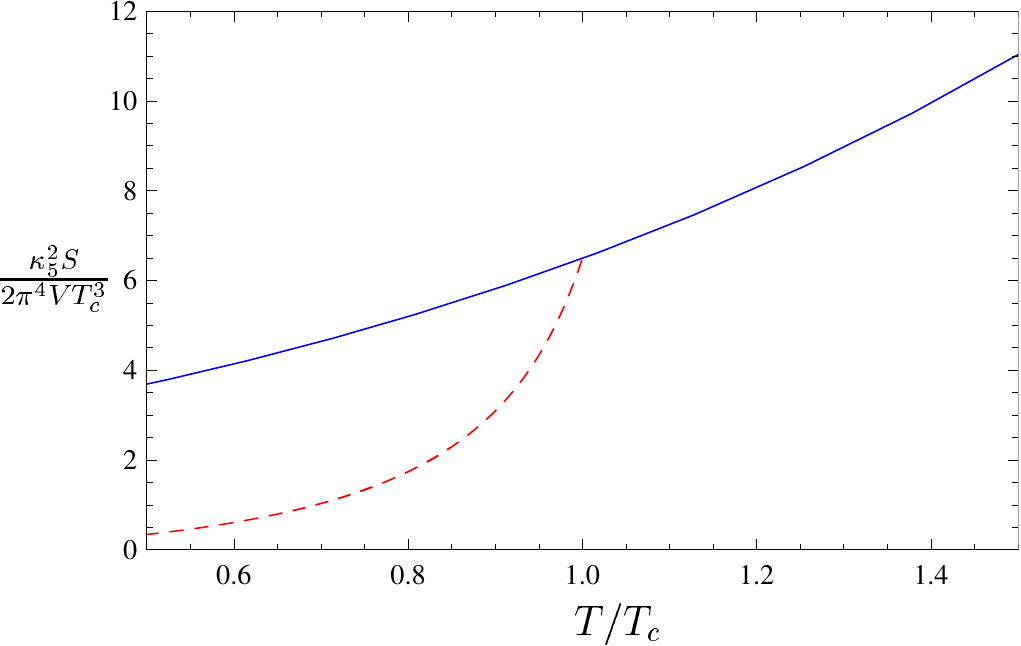}}
  \caption{(a) $\k^2\Omega/\left(\pi^4 V T_c^4\right)$ versus the rescaled
    temperature $T/T_c$ for $\alpha=0.316$. The blue solid curve comes from
    solutions with $\langle J^x_1 \rangle =0$ while the red dashed curve comes from solutions with nonzero $\langle J^x_1\rangle$. For $T>T_c$, we have only the blue curve, but when $T\leq T_c$ the red dashed curve appears and has the lower $\k^2\Omega/\left(\pi^4 V T_c^4\right)$, indicating a phase transition at $T=T_c$. $\k^2\Omega/\left(\pi^4 V T_c^4\right)$ is continuous and differentiable at $T=T_c$. (b) $\k^2S/\left(2\pi^4 V T_c^3\right)$ versus $T/T_c$ for $\alpha=0.316$. The blue solid and red dashed curves have the same meaning as in (a). $\k^2S/\left(2\pi^4 V T_c^3\right)$ is continuous but not differentiable at $T=T_c$, indicating a second-order transition.}
  \label{fig:thermoalpha0.1}
\end{figure}

For example, for $\alpha=0.316<\alpha_c$, we only find solutions with $\langle J^x_1 \rangle = 0$ until a temperature $T_c$ where a second set of solutions, with nonzero $\langle J^x_1\rangle$, appears. Figure~\ref{fig:cond} shows that $\langle J^x_1 \rangle$ rises continuously from zero as we decrease $T$ below $T_c$.
Figure~\ref{fig:thermoalpha0.1} (a) shows the grand potential $\Omega$,
divided by $\pi^4 V T_c^4/\k^2$, versus the rescaled temperature $T/T_c$ for
$\alpha=0.316$. The blue solid curve in Figure~\ref{fig:thermoalpha0.1} (a) comes from solutions with $\langle J^x_1 \rangle = 0$ and the red dashed curve comes from solutions with $\langle J^x_1 \rangle \neq 0$.
We see clearly that at $T<T_c$ the states with $\langle J^x_1 \rangle\neq0$ have the lower $\k^2\Omega/\left(\pi^4 V T_c^4\right)$ and hence are thermodynamically preferred. We thus conclude that a phase transition occurs at $T=T_c$. The nonzero $\langle J^x_1\rangle$ indicates spontaneous breaking of $U(1)_3$ and of $SO(3)$ rotational symmetry down to $SO(2)$, and hence is an order parameter for the transition. Figure~\ref{fig:thermoalpha0.1} (b) shows the entropy $S$, divided by $2\pi^4 V T_c^3/\k^2$, versus the rescaled temperature $T/T_c$ for $\alpha=0.316$. The blue solid curve and the red dashed curve have the same meaning as in Figure~\ref{fig:thermoalpha0.1} (a). Here we see that $\k^2S/\left(2\pi^4 V T_c^3\right)$ is continuous but has a kink, \textit{i.e.} a discontinuous first derivative, clearly indicating a second-order transition. For other values of $\alpha<\alpha_c$, the figures are qualitatively similar.

A good question concerning these second-order transitions is: what are the critical
exponents? In the probe limit, $\alpha=0$, an analytic solution for the gauge
fields exists for $T$ near $T_c$ \cite{Basu:2008bh}, which was used in ref.
\cite{Herzog:2009ci} to show that for $T \lesssim T_c$, $\langle J^x_1 \rangle
\propto \left( 1- T/T_c \right)^{1/2}$. In other words, in the probe limit the
critical exponent for $\langle J^x_1\rangle$ takes the mean-field value $1/2$.
Does increasing $\alpha$ change the critical exponent? Our numerical evidence
suggests that the answer is no: for all $\alpha < \alpha_c$, we appear to find
$\langle J^x_1 \rangle \propto \left( 1-T/T_c\right)^{1/2}$ (see Figure~\ref{fig:cond}).

\begin{figure}
  \centering
  \subfigure[]{\includegraphics[width=0.47\linewidth]{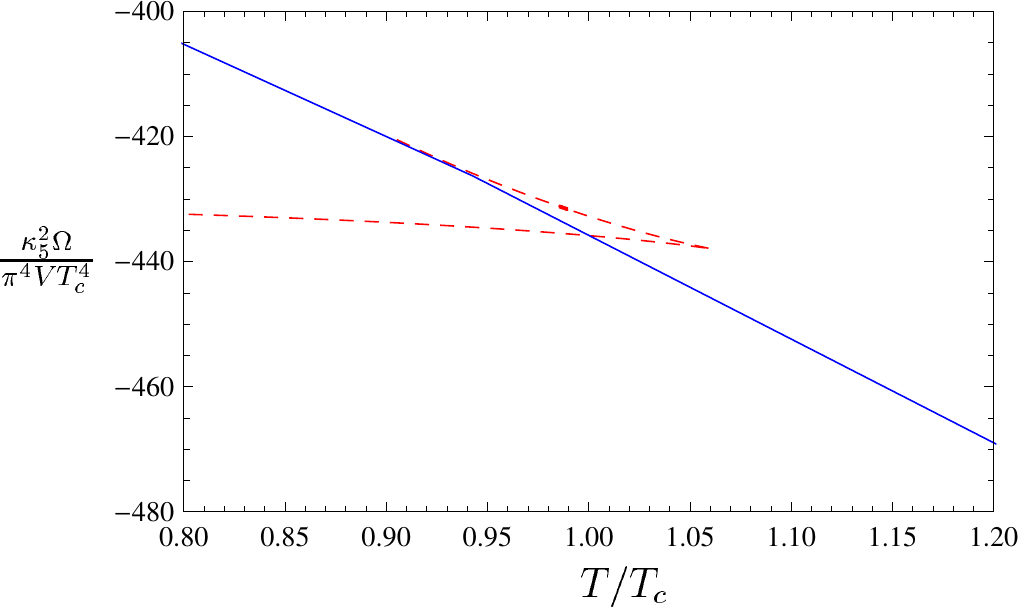}}
  \hfill
  \subfigure[]{\includegraphics[width=0.47\linewidth]{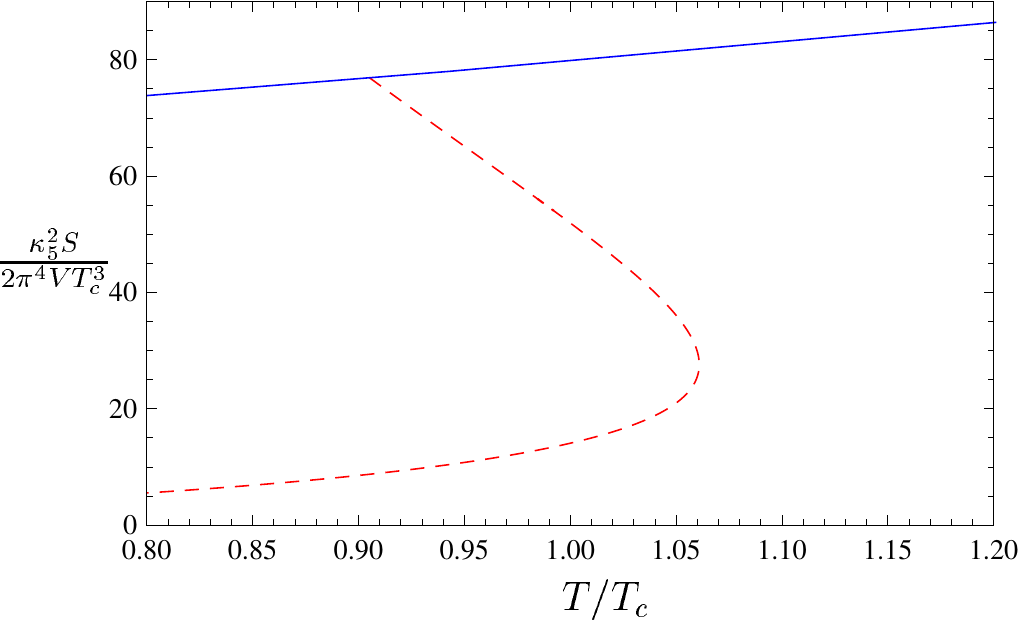}}
  \caption{(a) $\k^2\Omega/\left(\pi^4 V T_c^4\right)$ versus the rescaled temperature $T/T_c$ for $\alpha=0.447$. The blue solid and red dashed curves have the same meanings as in Figure~\ref{fig:thermoalpha0.1}. For $T>T_c$ we have only the blue solid curve. At $T=1.061\,T_c$, the red dashed curve appears and $\k^2\Omega/\left(\pi^4 V T_c^4\right)$ becomes multi-valued. When $T\leq T_c$ the red dashed curve has the lowest $\k^2\Omega/\left(\pi^4 V T_c^4\right)$, indicating a phase transition at $T=T_c$. $\k^2\Omega/\left(\pi^4 V T_c^4\right)$ is continuous but not differentiable at $T=T_c$, signaling a first-order transition. (b) $\k^2S/\left(2\pi^4 V T_c^3\right)$ versus $T/T_c$ for $\alpha=0.447$. $\k^2S/\left(2\pi^4 V T_c^3\right)$ is not continuous at $T=T_c$, but rather jumps from the blue solid curve to the lowest branch of the red dashed curve, indicating a first-order transition.} 
  \label{fig:thermoalpha0.2}
\end{figure}

As $\alpha$ increases past $\alpha_c=0.365\pm0.001$, we see a qualitative change in the thermodynamics. Consider for example $\alpha = 0.447$. Here again we only find solutions with $\langle J^x_1 \rangle=0$ down to some temperature where \textit{two} new sets of solutions appear, both with nonzero $\langle J^x_1\rangle$.  In other words, three states are available to the system: one with $\langle J^x_1\rangle=0$ and two with nonzero $\langle J^x_1 \rangle$. Figure~\ref{fig:cond} shows that as we cool the system, $\langle J^x_1 \rangle$ becomes multi-valued at $T=1.061 \, T_c$. To determine which state is thermodynamically preferred, we compute the grand potential $\Omega$. Figure~\ref{fig:thermoalpha0.2} (a) shows $\k^2\Omega/\left(\pi^4 V T_c^4\right)$ versus $T/T_c$. The blue solid curve and the red dashed curve have the same meanings as in Figure~\ref{fig:thermoalpha0.1}.
We immediately see the characteristic ``swallowtail'' shape of a first-order phase transition. If we decrease $T$, entering the figure along the blue solid curve from the right, we reach the temperature $T=1.061\,T_c$ where the new solutions appear (as the red dashed curve). The blue solid curve still has the lowest $\k^2\Omega/\left(\pi^4 V T_c^4\right)$ until $T=T_c$ (by definition). If we continue reducing $T$ below $T_c$, then the red curve has the lowest $\k^2\Omega/\left(\pi^4 V T_c^4\right)$. The transition is clearly first order: $\k^2\Omega/\left(\pi^4 V T_c^4\right)$ has a kink at $T=T_c$. We can also see from the entropy that the transition is first order. Figure~\ref{fig:thermoalpha0.2} (b) shows $\k^2S/\left(2\pi^4 V T_c^3\right)$ versus $T/T_c$. The entropy, like the grand potential, is multi-valued, and jumps discontinuously from the blue solid curve to the lowest part of the red dashed curve at $T=T_c$, indicating a first-order transition.

Notice that a crucial difference between $\alpha<\alpha_c$ (second order) and $\alpha>\alpha_c$ (first order) is that for $\alpha>\alpha_c$ the critical temperature $T_c$ is not simply the temperature at which $\langle J^x_1\rangle$ becomes nonzero. We need more information to determine $T_c$ when $\alpha>\alpha_c$, for example we can study $\Omega$.

A good question is: how does increasing $\alpha$ change $T_c$? Table~\ref{table:alphatc} shows several values of $\alpha$ and the associated $T_c$ in units of fixed $\mu$. In the probe limit, $\alpha=0$, we have the analytic result from ref. \cite{Herzog:2009ci} that $T_c/\mu = 1/4\pi \approx 7.96 \times 10^{-2}$. For finite $\alpha$, the general trend is that $T_c$ decreases as we increase $\alpha$.

\begin{table}[h!]
\begin{center}
\begin{tabular}{|c|c|}
\hline
$\alpha$ & $T_c/\mu \times 10^2$ \\ \hline \hline
$0$ & $7.96$ \\ \hline
$0.032$ & $7.92$ \\ \hline
$0.316$ & $4.57$ \\ \hline
$0.364$ & $3.62$ \\ \hline
$0.447$ & $2.18$ \\ \hline
\end{tabular}
\end{center}
\caption{Values of $T_c/\mu$, scaled up by a factor of $10^2$, for various values of $\alpha$. Recall that the critical value of $\alpha$ is $\alpha_c = 0.365\pm0.001$.}\label{table:alphatc}
\end{table}

%%%%%%%%%%%%%%%%%%%%%%%%%%%%%%%%%%%%%%%%%%%%%%%%%%%%%%%%%%%%%%%%%%%%%%%%%%%%%%%%%%%%%%%%%%%%%%%%%%%%%%%%%%%%%%
\section{Discussion and Outlook}
\label{discuss}
%%%%%%%%%%%%%%%%%%%%%%%%%%%%%%%%%%%%%%%%%%%%%%%%%%%%%%%%%%%%%%%%%%%%%%%%%%%%%%%%%%%%%%%%%%%%%%%%%%%%%%%%%%%%%%

We studied asymptotically AdS charged black holes  in $(4+1)$-dimensional $SU(2)$ Einstein-Yang-Mills theory with finite $\alpha = \k/\hat{g}$, that is, with back-reaction of the gauge fields. Our numerical solutions show that, for a given value of $\alpha$, as the temperature decreases the black holes grow vector hair. Via AdS/CFT, this process appears as a phase transition to a p-wave superfluid state in a strongly-coupled CFT. We have shown that the order of the phase transition depends on the value of $\alpha$: for values below $\alpha_c=0.365\pm0.001$, the transition is second order, while for larger values the transition is first order.

As we mentioned in the introduction, intuitively we may think of increasing $\alpha$ as increasing the ratio of charged degrees of freedom to total degrees of freedom in the CFT. To make that intuition precise, we can consider a specific system. One string theory realization of $SU(2)$ gauge fields in AdS space is type IIB supergravity in $(4+1)$-dimensional AdS space (times a five-sphere) plus two coincident D7-branes that provide the $SU(2)$ gauge fields \cite{Ammon:2008fc,Basu:2008bh,Ammon:2009fe,Peeters:2009sr}. The dual field theory is $\N=4$ supersymmetric $SU(N_c)$ Yang-Mills theory, in the limits of large $N_c$ and large 't Hooft coupling, coupled to a number $N_f=2$ of massless $\N=2$ supersymmetric hypermultiplets in the $N_c$ representation of $SU(N_c)$, \textit{i.e.} flavor fields. The global $SU(N_f)=SU(2)$ is an isospin symmetry. Translating from gravity to field theory quantities, we have $1/\k^2\propto N_c^2$ and $1/\hat{g}^2\propto N_fN_c$, hence $\alpha \propto \sqrt{N_f/N_c}$, which supports our intuition. We must be cautious, however. In the field theory, the probe limit consists of neglecting quantum effects due to the flavor fields because these are suppressed by powers of $N_f/N_c$. If $N_f/N_c$ becomes finite, then, for example, in the field theory the coupling would run, the dual statement being that in type IIB supergravity the dilaton would run, which is an effect absent in our model. We should not draw too close an analogy between our simple model and this particular string theory system.

Returning to our simple system, with fully back-reacted solutions we can potentially answer many questions:

What happens as we increase $\alpha$ further? As $\alpha$ increases, $T_c$ appears to decrease. An obvious question is whether $T_c$ ever becomes zero. The second-order transition occurs because of an instability of the gauge field in the Reissner-Nordstr\"om background \cite{Gubser:2008wv}. At $T=0$, that instability only exists for values of $\alpha$ below some upper bound \cite{Basu:2009vv}. If the transition was always second-order, then we would conclude that the transition is only possible for $\alpha$ below the bound: if $\alpha$ is greater than the bound, then the instability never appears, even if we cool the system to $T=0$. If the transition becomes first-order, however, then we must rethink the bound: now Reissner-Nordstr\"om becomes metastable, and a phase transition occurs, at temperatures above those where the instability appears, so now $T_c$ may go to zero for some $\alpha$ above the bound (if at all).

Does a new scaling symmetry emerge at zero temperature? The zero-temperature analysis of ref. \cite{Basu:2009vv} suggests that the metric exhibits a new, emergent, scaling symmetry. Our preliminary numerical results suggest that indeed, in the approach to zero temperature, an emergent scaling symmetry appears, of the form suggested in ref. \cite{Basu:2009vv}. The $T\rightarrow0$ limit is numerically challenging, however, so a firm answer must wait.

What about the transport properties of the dual conformal fluid, for example the electrical conductivity, which at zero temperature should exhibit a ``hard gap,'' as explained in ref. \cite{Basu:2009vv}? What is the fluid's response to nonzero superfluid velocities? In similar systems, sufficiently large superfluid velocities also changed the transition from second to first order \cite{Basu:2008st,Herzog:2008he}. What is the speed of sound, which need not be the same in all directions since rotational symmetry is broken, or the speeds of second and fourth sounds \cite{Herzog:2009ci,Yarom:2009uq,Herzog:2009md}?

We plan to investigate these and related questions in the future.

%%%%%%%%%%%%%%%%%%%%%%%% A C K N O W L E D G M E N T S
\paragraph{ACKNOWLEDGEMENTS}
\addcontentsline{toc}{section}{Acknowledgments}
We are grateful to A. Buchel and M. Rangamani for discussions. This work was supported in part by  {\it The Cluster of Excellence for Fundamental Physics - Origin and Structure of the Universe}. M. A. would also like to thank the Studienstiftung des deutschen Volkes for financial support.

%%%%%%%%%%%%%%%%%%%%%%%% R E F E R E N C E S

\providecommand{\href}[2]{#2}\begingroup\raggedright\endgroup

%\bibliographystyle{felice_utcaps}
%\bibliography{/home/pcl247e/pkerner/texmf/bibtex/bib/papers,/home/pcl247e/pkerner/texmf/bibtex/bib/books,/home/pcl247e/pkerner/texmf/bibtex/bib/papersunpub,/home/pcl304/jke/Matthias/pwavebackreact/letter/temp} 

\end{document}